\documentclass[12pt,preprint]{article}%
\usepackage{amssymb}
\usepackage{amsmath}
\usepackage{graphics}
\usepackage{epsfig}
\usepackage{amsfonts}
\usepackage{graphicx}%
\newcommand{\eqb}{\begin{equation}}
\newcommand{\eqe}{\end{equation}}
\newcommand{\dmb}{\begin{displaymath}}
\newcommand{\dme}{\end{displaymath}}

\newcommand{\eab}{\begin{eqnarray}}
\newcommand{\eae}{\end{eqnarray}}

\newcommand{\be}{\begin{equation}}
\newcommand{\ee}{\end{equation}}

\begin{document}
\begin{titlepage}
\begin{center}
\Large{The isolated electron: \\ De Broglie's "hidden" thermodynamics, SU(2) Quantum Yang-Mills theory, 
and a strongly perturbed BPS monopole}\vspace{1.5cm}\\ 
\large{Ralf Hofmann}
\end{center}
\vspace{1.5cm}
\begin{center}
{\em Institut f\"ur Theoretische Physik\\
Universit\"at Heidelberg\\
Philosophenweg 16\\
69120 Heidelberg, Germany}
\end{center}
\vspace{0.2cm}
\begin{center}
{and}
\end{center}
\vspace{0.2cm}
\begin{center}
{\em Institut f\"ur Photonenforschung und Synchrotronstrahlung\\  
Karlsruher Institut f\"ur Technologie\\ 
Hermann-von-Helmholtz-Platz 1\\ 
D-76344 Eggenstein-Leopoldshafen, Germany}
\end{center}
\vspace{1.5cm}

\begin{abstract}
Based on a recent numerical simulation of the temporal evolution 
of a spherically perturbed BPS monopole, SU(2) Yang-Mills thermodynamics, 
Louis de Broglie's deliberations on the disparate Lorentz transformations of the frequency of an internal "clock" on one hand and 
the associated quantum energy on the other hand, and postulating that the electron is represented by a figure-eight shaped, self-intersecting center vortex loop in SU(2) Quantum Yang-Mills theory we estimate the spatial radius $R_0$ of this self-intersection region in terms 
of the electron's Compton wave length $\lambda_C$. This region, which is 
immersed into the confining phase, constitutes a blob of deconfining 
phase of temperature $T_0$ mildly above the critical temperature $T_c$ carrying a frequently perturbed BPS monopole 
(with a magnetic-electric dual interpretation of its charge w.r.t. U(1)$\subset$SU(2)). We also establish a quantitative relation between 
rest mass $m_0$ of the electron and SU(2) Yang-Mills scale $\Lambda$, which in turn 
is defined via $T_c$. Surprisingly, $R_0$ turns out to 
be comparable to the Bohr radius while the core size of the monopole matches $\lambda_C$, and the correction to 
the mass of the electron due to Coulomb energy is about 2\,\%. 
\end{abstract}
\end{titlepage}

\section{Introduction}

The electron and other charged leptons are considered fundamental particles in the present Standard Model 
of particle physics. According to the overwhelming majority of data on (high-energy) electron-electron/electron-positron collisions and atomic physics such a classification rests on the fact that no structural charge-density and spin distributions  are revealed by typical scattering processes and spectroscopic experiments. 
Moreover, in many condensed-matter systems the notion of the Quantum Mechanics (QM) of a point-like, spinning 
particle together with the exclusion principle do describe the collective behavior of electrons in a 
highly satisfactory and realistic way \cite{densityfunctional}. However, 
the quantitative description of strongly 
correlated two-dimensional electrons associated, e.g., with high-temperature 
superconductivitiy, seems to demand a deviating treatment. Here a spatial separation between 
charge and spin is a serious option favored by the data \cite{Anderson}. 

To describe the spatial probability 
density for the presence of a point-like electron in terms of the "square" of a 
wave function based on de Broglie's particle-wave duality \cite{deBroglieDoc}, 
whose time evolution is governed by the Hamiltonian of the isolated system, is an 
extremely successful and useful concept: About a century ago, it started to revolutionize 
our understanding of atomic stability and of the discreteness of the spectra 
of light emitted by excited atoms \cite{Bohr,Sommerfeld,Heisenberg,Schroedinger,Dirac}, the chemical bond \cite{Pauling}, and 
the role of and interplay between electrons in condensed matter and 
hot plasmas thanks to the development of efficient calculational procedures for 
(extended) multi-electron systems. Moreover, new computational and conceptual horizons, enabled by 
quantum field theoretical (QFT) second quantization in the framework of Quantum Electrodynamics (QED) \cite{Tomonoga,Feynman,Schwinger} and its electroweak generalization \cite{Lee&Yang,Glashow,Salam,Weinberg}, have  
yielded accurate predictions of radiative corrections to atomic 
energy levels, to the electron's magnetic moment, and in electroweak scattering cross 
sections. 

Why then is there a need for a deeper understanding 
of the nature of the free electron and its properties as revealed by the application of external forces, the implied 
(classical) radiation reaction, and tree-level as well as radiative quantum behavior playing 
out in scattering processes? The answer to this question touches a number of basic problems. 

First, already on the classical level Maxwell's equation and the Lorentz force 
equation do not describe the phenomenon of radiation reaction: This local field theory ignores the back-reaction onto an electron induced by the radiation emitted by this particle when under acceleration in an external field. Various 
proposals on how to modify the force equation were made in the literature, see, e.g., \cite{Hadad,HadadRafelski}, 
but a basic theory yielding an effective, unique correction is not 
available. As was shown in \cite{HadadRafelski}, the effects of radiation reaction can dominate 
the dynamics of the electron for strong accelerations posited by high-power laser pulses.     

Second, even though QM and QFT provide an efficient and reliable computational 
framework there are conceptual questions. Why does the electron as a particle have a modest 
rest mass of 
\eqb
\label{measuredmass}
m_0=0.511\,\mbox{MeV/$c^2$}
\eqe
if probing its structure by high-energetic scattering experiments at four-momentum transfers $Q$ up 
to several hundreds of GeV/$c$ does not indicate any deviation 
from structurelessness or point-particle behavior? Classically speaking, such a point-likeness would imply the spatial integral of its 
Coulombic electric field energy density and hence its rest mass $m_0$
to be of order $Q$. Namely, in natural units $c=\hbar=1$ one has 
\eqb\label{masselec}
m_0\sim\int_{Q^{-1}}^{\infty} dr\,\frac{r^2}{r^4}=Q\,
\eqe
which contradicts the value in Eq.\,(\ref{measuredmass}). 
Notice that radiative corrections to $m_0$, computable in QED in terms of powers of a small coupling 
constant $\alpha\sim 1/137$, must, by the 
very definition of a renormalized perturbation series, be much smaller than $m_0$. Concerning another aspect of 
the electron's potential structure, one may ask how a finite magnetic moment can possibly be related to the spin of a point particle? 
Yet, although (or possibly because) it ignores this basic question, representation theory in QM is immensely successful in classifying the effect of coupled angular momentum on energy levels and on the overall dynamics of composite systems.

\begin{figure}[h]
  \centerline{\includegraphics[width=200pt]{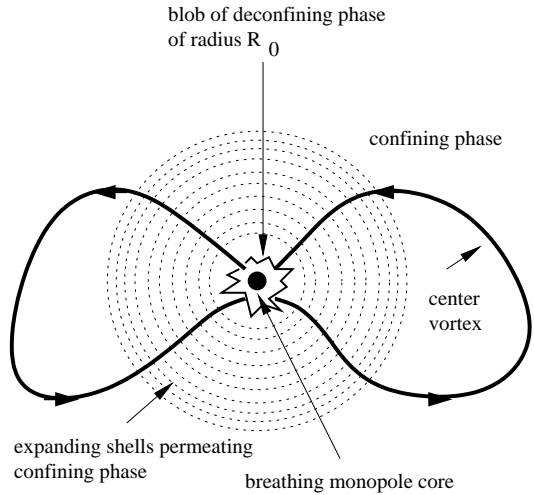}}
  \caption{Schematics of a one-fold self-intersecting center-vortex loop, immersed 
  into the confining phase of SU(2) Yang-Mills theory. SU(2) Yang-Mills theory is to be interpreted in an electric-magnetically dual way, and therefore the breathing core of the perturbed BPS monopole, in turn immersed into the (fuzzy) intersection region of radius $R_0$ endowed with deconfining energy density $\rho$, actually represents 
  an electric charge. Also, the magnetic center flux, residing in the two wings of the vortex-loop,  
  actually is an electric one, giving rise to a magnetic moment twice that of a 
  single vortex loop. Spin-1/2 is represented by the two 
  possible directions of (dually interpreted) ${\bf Z}_2$ center flux. The associated 
  local ${\bf Z}_2$ symmetry is broken dynamically in the confining phase by the 't Hooft loop acquiring a finite expectation \cite{HofmannBook2016}. 
  The concentric circles indicate the dilution 
  of the confining phase by expanding high-frequency shells carried by 
  massive off-Cartan modes of the monopole. These shells are induced by the action of 
  (anti)caloron centers perturbing the BPS monopole.      \label{fig-1}}
\end{figure}

Third, Louis de Broglie's deep ideas, 
underlying the proposal of 
wave-particle duality for the electron \cite{deBroglieDoc}, in their original form imply that the electron is 
anything but a particle of vanishing spatial extent. Strictly speaking, 
this contradicts Born's intepretation of the square of the wave function describing the probability density for the spatial occurrence of a point particle. De Broglie argues \cite{DeBroglieRev} by considering respective changes, under a Lorentz boost at velocity $v$, of the electron's rest mass $m_0$, associated with the frequency $\nu_0$ of an internal "clock" oscillation 
via Planck's quantum of action $h$ as  
\eqb
\label{mass}
m_0c^2=h\nu_0\,,
\eqe
and viewed as the zero component of the electron's four-momentum. Subsequently, he contrasts this with the 
disparate changes of the same {\sl internal} frequency as implied by time dilatation. As a consequence, 
the increase of particle energy from $m_0c^2$ to 
$mc^2=\frac{m_0c^2}{\sqrt{1-\beta^2}}$ can be decomposed into a reduction of 
internal heat from $m_0c^2\equiv Q_0$ to $Q=Q_0\sqrt{1-\beta^2}$ plus an increase of (quasi-)translational energy from zero to $vp$:
\eqb
\label{deBrogliedeco}
mc^2=\frac{m_0c^2}{\sqrt{1-\beta^2}}=Q+vp\equiv m_0c^2\sqrt{1-\beta^2}+vp\,,
\eqe 
where $\beta\equiv v/c$, the relativistic spatial momentum is given 
as $p\equiv \frac{m_0 v}{\sqrt{1-\beta^2}}$, and $c$ denotes 
the speed of light in vacuum. Notice that $Q=Q_0\sqrt{1-\beta^2}$ refers to Planck's formulation of relativistic 
thermodynamics which, in a straight forward way and without directly addressing dissipative processes,  
assures proper Lorentz-transformation behavior of an exhaustive set of thermodynamical quantities. This may be contrasted with Ott's formulation \cite{Ott,TerHaarWergeland} whose justification appears to be rather mysterious to the present author. 
Notice also that the second term on the right-hand side of 
Eq.\,(\ref{deBrogliedeco}) -- the (quasi-)translational energy -- reduces to nonrelativistic 
kinetic energy $\frac12 m_0 v^2$ only modulo a factor of two for $v\ll c$. 
This is an unexplained point which eventually needs to be clarified. It is absurd, however, to assign 
an internal heat $Q$ to a point particle. The present work proposes that the thermodynamics of an isolated electron at 
rest and the existence of a preferred internal frequency $\nu_0=\frac{m_0c^2}{h}$ are consequences of the existence of a compact, spatial region containing an electric, BPS-like monopole (after an electric-magnetically dual interpretation \cite{HofmannBook2016}), which accommodates the electron's charge and mass, representing a blob of deconfining phase of SU(2) Quantum Yang-Mills theory immersed into a confining-phase environment. Macroscopically seen, this blob represents the self-intersection of an SU(2) center vortex loop, that is, 
(again, after an electric-magnetically dual interpretation \cite{HofmannBook2016}) a figure-eight 
shaped, one-fold knotted loop of electric center flux inducing a magnetic moment 
twice as large as the fundamental unit provided by a single center-vortex loop, see Fig.\,\ref{fig-1}. 

Finally, deeper insight 
into the functioning of the electron as a quantum particle is destined to shed 
light on the "sweet mysteries" of Quantum Mechanics (e.g. instantaneous state reduction by measurement) 
and the renormalization problem in 
perturbative Quantum Field Theory.       

The present work offers a conceptual scheme for the description of the unaccelerated 
electron which supports de Broglie's ideas. This scheme draws on the following developments: 
(i) Phase structure of SU(2) Yang-Mills thermodynamics \cite{HofmannBook2016}, composition and 
wave/particle like excitability of the deconfining thermal ground state \cite{Hofmann2016,GrandouHofmann2015}, 
and principle nature of solitonic excitations in the confining phase 
\cite{MoosmannHofmannI,MoosmannHofmannII}. 
(ii) Recent results on the classical dynamics of the strongly (and isotropically) perturbed 
BPS monopole \cite{FodorRacz2004}. Pairing with the work in \cite{FodorRacz2004}, 
there are insightful and supporting treatments of linear perturbations of the BPS monopole (spectrum of quasi-normal modes in \cite{Forgacs}). 

As a consequence of (i) and (ii) the quantum thermodynamics within the self-intersection region of a center-vortex loop 
manifests itself as follows (in natural units): Created by the dissociation of a large-holonomy caloron, a BPS monopole, with the asymptotic value of its Higgs field modulus at temperature $T_0$ given as
\eqb
\label{Higgsasy}
H_{\infty}(T_0)=\pi T_0\ 
\eqe
due to maximum (anti)caloron holonomy\footnote{Temperature $T_0$ of the 
intersection region is slightly higher than the critical temperature $T_c$ of the deconfining-preconfining phase transition: $T_0=1.315\,T_c$, see Sec.\,\ref{decthermgs}.} \cite{Nahm,KraanVanBaal,LeeLu,Diakonov2004} and the value of the coupling constant $e$ 
in the adjoint Higgs model roughly determined by the plateau value $e=\sqrt{8}\pi$ \cite{HofmannBook2016}, 
is immersed into deconfining SU(2) Yang-Mills thermodynamics. At $T_0=1.315\,T_c$, see Sec.\,\ref{PSE}, the pressure of deconfining SU(2) Yang-Mills thermodynamics vanishes. The contribution to the total pressure from 
a static (noninteracting) BPS monopole \cite{'tHooftMonop,BPS} is also nil. In the absence of interactions between such an explicit monopole and deconfining SU(2) Yang-Mills thermodynamics the intersection region would thus be static at temperature $T_0$, that is, non-expanding and non-contracting.   

The unperturbed monopole's rest mass is given as \cite{'tHooftMonop,BPS}
\eqb
\label{monopolemass}
m_{\rm mon,0}=\frac{8\pi^2}{e^2}H_{\infty}\sim H_{\infty}\,,
\eqe
where $H_{\infty}$ denotes its asymptotic adjoint Higgs field, and the energy density $\rho$ of deconfining SU(2) Yang-Mills thermodynamics at 
$T_0$ reads 
\eqb
\label{rhogs}
\rho(T_0)=8.31\,\rho^{\rm gs}(T_0)\,,
\eqe
see Sec.\,\ref{PSE}, where $\rho^{\rm gs}=4\pi\Lambda^3 T$ denotes the energy density of the 
thermal ground state, and $\Lambda$ is the Yang-Mills scale, related to $T_c$ as
\eqb
\label{TCLambda}
T_c=\frac{13.87}{2\pi}\,\Lambda=\frac{1}{1.32} T_0\,.
\eqe
 
The fact that the explicit monopole is subject to perturbations, issued by 
quantum kicks due to caloron/anticaloron centers \cite{Hofmann2016}, leads to a fluctuating re-distribution 
of stress-energy within the self-intersection region. This situation is characterized, e.g., by the existence of 
a breathing mode of the monopole core whose (circular) frequency $\omega_0$ essentially represents $m_0$ \cite{DeBroglieRev} 
and is given by the mass $m_w=eH_\infty$ of the vector modes \cite{FodorRacz2004}.

The remainder of this paper is intended to corroborate and explain the above-sketched model of the electron 
in more quantitative terms, thereby considering a few (rather small) 
corrections to the results obtained in the simplified treatment of \cite{Hofmann2017} 
where complete ground-state dominance of the deconfining thermodynamics within 
the intersection region was assumed. Since our derivations heavily rest on a grasp 
of the phase structure of an SU(2) Yang-Mills theory we will briefly review it in 
Sec.\,\ref{PSE}. Sec.\,\ref{decthermgs} discusses the deconfining pressure $P$ and derives the 
ratio $\rho(T_0)/\rho^{\rm gs}(T_0)$ and the value of the coupling $e$ (slightly higher than $\sqrt{8}\pi$) 
at the zero $T_0$ of $P$. The mass of the noninteracting BPS monopole, 
given by the first of Eqs.\,(\ref{monopolemass}), thus is slightly decreased compared to the value given 
by the expression to the far right, and the mass $m_w$ of the vector modes is mildly increased. The results obtained in 
\cite{FodorRacz2004} on the dynamics of the spherically perturbed monopole are reviewed 
in Sec.\,\ref{Fodor}. Finally, in Sec.\,\ref{IM} we combine the results of 
Secs.\,\ref{PSE}, \ref{decthermgs}, and \ref{Fodor} to propose a 
model for the electron as a genuine quantum particle of finite 
extent in Sec.\,\ref{IM}. Sec.\,\ref{Sum} summarizes this work and gives an 
outlook to future activity.

\section{Review on phase structure of SU(2) Yang-Mills thermodynamics\label{PSE}}

SU(2) Yang-Mills thermodynamics occurs in three distinct phases \cite{Hofmann2005}. 
The high-temperature, deconfining phase is characterized by a 
thermal ground state, composed of overlapping, topological-charge-modulus 
unity calorons and anticalorons. At a given temperature $T\ge T_c$, the spatial extent of their densely 
packed centers is determined by a radius $\rho_u=|\phi|^{-1}=\sqrt{2\pi T/\Lambda^3}$ 
which also sets the preferred (anti)caloron scale parameter for the spatial coarse-graining 
process involving a homogeneously and adjointly transforming 
two-point function of the fundamental Yang-Mills field-strength 
tensor $F_{\mu\nu}$ evaluated on a Harrington-Shepard (HS) (that is, trivial-holonomy) 
caloron and its anticaloron \cite{HS1977,HerbstHofmann2004}. Here $|\phi|$ denotes the modulus 
of the emergent, adjoint, and inert scalar field representing coarse-grained 
(anti)caloron centers, $T$ refers to temperature as defined by the inverse 
temporal period of the HS-(anti)caloron, and $\Lambda$ is the Yang-Mills 
scale. The latter emerges as an integration constant when solving a first-order, 
ordinary differential equation for $\phi$'s potential. This first-order equation 
expresses the intactness of (Euclidean) BPS saturation within 
(anti)caloron centers. A departure from (anti)selfduality in the field 
configuration, reflecting the overlap of (anti)caloron peripheries, is manifested 
by a finite ground-state pressure $P^{\rm gs}$ and energy 
density $\rho^{\rm gs}=-P^{\rm gs}=4\pi\Lambda^3 T$. The associated, effective 
gauge-field configuration $a^{\rm gs}_\mu$ is obtained as a zero-curvature solution to 
the effective Yang-Mills equation $D_\mu G_{\mu\nu}=2ie[\phi,D_\nu\phi]$ where 
$G_{\mu\nu}\equiv\partial_\mu a_\nu-\partial_\nu a_\mu+ie[a_\mu,a_\nu]$ denotes the 
effective Yang-Mills field-strength tensor. 

Excitations of the deconfining thermal ground state are partially massive (mass $m=2e|\phi|$)
by virtue of the adjoint Higgs mechanisam invoked by the field $\phi$. 
These massive excitations never propagate in a wavelike 
way: their high would-be frequencies probe the interior of (anti)caloron 
centers to provoke an indeterministic, quantum-like response. On the level of free thermal quasi-particles (that is, not taking into account feeble radiative effects \cite{BischerGrandouHofmann2017,HofmannBook2016}) the massive sector thus 
merely represents uncorrelated, Bose-Einstein distributed 
energy and momentum fluctuations. The {\sl massless} sector, on the other hand, may propagate in a wavelike 
fashion if a constraint on intensity and frequency is satisfied \cite{Hofmann2016}, 
governed by the Yang-Mills scale $\Lambda$ of the theory.    

The thermal ground state of the deconfining phase rearranges itself into a different structure 
at the critical temperature $T_c=\frac{13.87}{2\pi}\,\Lambda$ where 
screened BPS monopoles and antimonopoles, 
released by rarely occurring large-holonomy calorons and anticalorons, become massless, therefore 
abundant, and, as a consequence, form a condensate. This marks the onset of the preconfining phase. 
Energetically, the condensate of monopoles and antimonopoles together 
with its massive gauge-mode excitations (Meissner-Ochsenfeld effect or Abelian Higgs mechanism \cite{PhilAnderson}) is 
not stable immediately at $T_c$ since the energy density of the deconfining phase is lower 
within the following temperature range \cite{HofmannBook2016}:
\eqb
\label{Tstar}
T_*\equiv 0.88 T_c\le T\le T_c\,.
\eqe

Shortly below $T_*$ the entropy density of the system approaches zero, and the thermal ground 
state of the preconfining phase decays into (spatial) $n$-fold self-intersecting 
center-vortex loops ($n=0,1,2,\cdots$). Since their masses 
scale as $n\Lambda$ but their multiplicities (number of distinct topologies at given $n$) 
scale more than factorially in $n$ \cite{LowTYM,BenderWu} the very concept of a partition function is 
inapplicable. This is the characteristics of a highly nonthermal 
Hagedorn transition: the homogeneity of pressure, energy density, and 
other "would-be" thermodynamical quantities is strongly violated, and the SU(2) 
Yang-Mills plasma exhibits a highly turbulent behavior.

\section{Pressureless deconfining SU(2) thermodynamics at $T_0$\label{decthermgs}}

\begin{figure}[h]
  \centerline{\includegraphics[width=200pt]{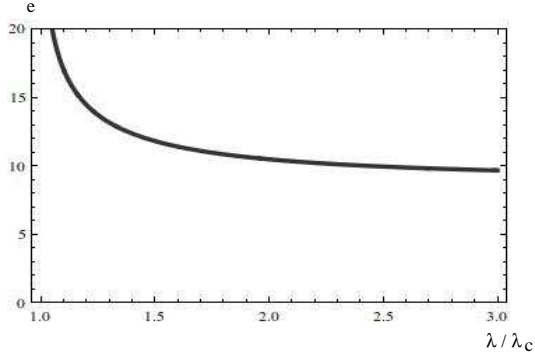}}
  \caption{Evolution of the coupling $e$ in the deconfining phase of SU(2) Yang-Mills thermodynamics as a function of $1\le\lambda/\lambda_c\le 3$. At $\lambda\sim\lambda_c=13.87$ the theory undergoes a second-order like phase transition towards the preconfining phase which is signalled by a (logarithmically thin) divergence of $e$.}\label{fig0}
\end{figure}

For accuracies on the 1\% level 
it is sufficient to consider thermodynamical quantities in 
one-loop approximation. Thermodynamical consistency up to this loop order implies an evolution equation 
for the coupling $e$ \cite{HofmannBook2016}. Its solution is depicted in Fig.\,\ref{fig0} as a function of $\lambda/\lambda_c$ where $\lambda\equiv\frac{2\pi T}{\Lambda}$ and $\lambda_c=13.87$. 

\begin{figure}[h]
  \centerline{\includegraphics[width=500pt]{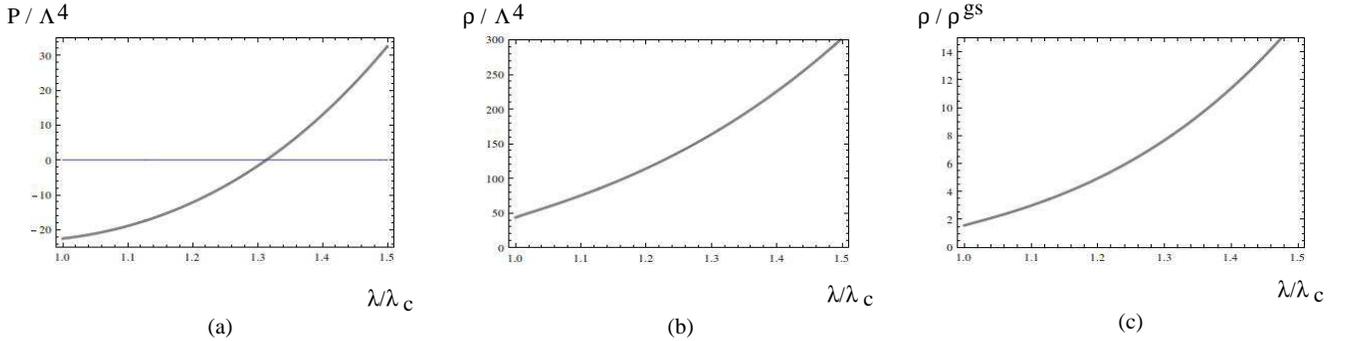}}
  \caption{Thermodynamical quantities of deconfining SU(2) Yang-Mills thermodynamics close to the deconfining-preconfining phase transition. (a): dimensionless pressure $P/\Lambda^4$, (b): dimensionless energy density 
  $\rho/\Lambda^4$, and (c): $\rho/\rho^{\rm gs}$ all as functions of $\frac{\lambda}{\lambda_c}$. At  $\lambda\sim\lambda_c=13.87$ the theory undergoes a second-order like phase transition towards the preconfining phase.}
\label{fig1}
\end{figure}

The deconfining-phase pressure $P$ and energy density $\rho$ are then 
given as \cite{HofmannBook2016}
\eab
\label{Ponelooprhooneloop}
P(\lambda)&=&-\Lambda^4\left\{\frac{2\lambda^4}{(2\pi)^6}\left[2\bar{P}(0)+6\,\bar{P}(2a)\right]+2\lambda\right\}\,,\nonumber\\ 
\rho(\lambda)&=&\Lambda^4\left\{\frac{2\lambda^4}{(2\pi)^6}\left[2\bar{\rho}(0)+6\,\bar{\rho}(2a)\right]+2\lambda\right\}\,,
\eae
where 
%********
\eab
\label{defdimPdimrho}
\bar{P}(y)&\equiv&\int_0^\infty
dx\,x^2\,\log\left[1-\exp(-\sqrt{x^2+y^2})\right]\,,\nonumber\\  
\bar{\rho}(y)&\equiv&\int_0^\infty
dx\,x^2\frac{\sqrt{x^2+y^2}}{\exp(\sqrt{x^2+y^2})-1}\,,
\eae
%*******
%******
\eqb
\label{defamass}
a\equiv\frac{m}{2T}\,.
\eqe
%********
In Eq.\,(\ref{defamass}) $m=2e\sqrt{\frac{\Lambda^3}{2\pi T}}$ 
denotes the thermal quasi-particle 
mass of the vector modes (unitary gauge) as induced by the adjoint Higgs mechanism 
effectively inaugurated by (anti)caloron centers. In 
Fig.\,\ref{fig1} $P/\Lambda^4$, $\rho/\Lambda^4$, and $\rho/\rho^{\rm gs}$ are depicted for $\lambda_c=13.87\le\lambda\le 1.5\lambda_c$. 
Notice the zero of $P$ (positive pressure of (quasi-particle) excitations balanced by negative ground-state pressure) 
at 
\eqb
\label{zerolam}
\lambda_0=1.32\,\lambda_c=18.31
\eqe
in Fig.\,\ref{fig1}a where 
\eqb
\label{rhovsrhogsT0}
\rho/\rho^{\rm gs}(1.32)=8.31\,,
\eqe 
see Fig.\,\ref{fig1}c, meaning that (quasi-particle) excitations already dominate the energy density at $\lambda_0$. On the other hand, 
one has $\rho/\rho^{\rm gs}(1)=1.56$, indicating the importance 
of the ground state energy density at the deconfining-preconfining phase boundary, see Fig.\,\ref{fig1}c. 
Finally, at $\lambda_0$ the 
coupling is
\eqb
\label{ecout0}
e(1.32)=12.96\,.
\eqe
 
\section{Review on strongly perturbed BPS monopole\label{Fodor}}

The normal-mode spectrum of a 't Hooft-Polyakov monopole in the BPS limit (considering small field fluctuations only) 
was investigated in \cite{Forgacs}. In \cite{FodorRacz2004} a spherically symmetric, 
strong perturbation of the BPS monopole in the SU(2) Yang-Mills-adjoint-Higgs model 
was analysed numerically as a nonlinear initial-value problem 
by virtue of a hyperboloidal conformal transformation of the original field equations for the profiles $H(r,t)=h(r,t)/r+H_\infty$ and $w(r,t)$ of the adjoint Higgs and 
off-Cartan fields, respectively. Their results can be summarized as follows. Considering a spherically symmetric, 
localized initial pulse as a strong perturbation of the static BPS monopole, the typical 
dynamical response did not depend on the parameter values of this pulse within a wide range. 
There are high-frequency oscillations in $w$ which form expanding shells 
decaying in time as $t^{-1/2}$. (The further away from the monopole 
core the shell is the higher the frequencies are that build it.) Moreover, a localized 
breathing state appears in association with the energy density of the monopole core 
region whose frequency $\omega_0$ approaches the mass $m_w=eH_\infty$ ($e$ denoting the gauge coupling) 
of the two off-Cartan modes in a power-like way in time (natural units): 
\eqb
\label{freutomass}
\omega_0=eH_\infty-C_w t^{-2/3} \ \ \Rightarrow \ \ \lim_{t\to\infty}\omega_0=eH_\infty\,,
\eqe
where $C_w$ is a positive constant. The amplitude of 
the oscillation in energy density decays like $C_a t^{-5/6}$, $C_a$ again denoting a 
positive constant. Disregarding for now the question what the physics of the exciting initial condition 
is, an internal clock of (circular) frequency $\omega_0\sim m_w$ is therefore run within the core region 
of the perturbed monopole.

\section{Size estimates\label{IM}}

Based on  \cite{Hofmann2017} we would now like to perform an improved estimate of the radius 
$R_0$ associated with the region of self-intersection. The idea is to prescribe at 
$T_0$, where deconfining SU(2) thermodynamics does not exert any pressure, that the rest mass of the quantum particle 
electron,  
\eqb
\label{massele}
m_0\sim e(T_0) H_\infty(T_0)=12.96\,H_\infty(T_0)\,,
\eqe  
see Eqs.\,(\ref{freutomass}) and (\ref{ecout0}), decomposes into that of a static monopole 
\eqb
\label{monopolemassNew}
m_{\rm mon,0}=\frac{8\pi^2}{e^2(T_0)}(T_0)H_{\infty}(T_0)=\frac{8\pi^2}{12.96^2}H_{\infty}\,,
\eqe
compare with Eqs.\,(\ref{monopolemass}) and (\ref{ecout0}), and the energy $E_0$ of deconfining SU(2) thermodynamics 
contained in the volume $V=\frac43 R_0^3$:
\eab
\label{masseNew}
m_0&=&12.96\,H_\infty(T_0)=m_m+E_0=\frac{8\pi^2}{12.96^2}H_\infty(T_0)+\frac43\pi R_0^3\,\rho(T_0)\nonumber\\ 
&=&H_\infty(T_0)\left(\frac{8\pi^2}{12.96^2}+
8.31\times\frac{128\pi}{3} \left(\frac{R_0}{18.31}\right)^3 H^3_\infty(T_0)\right)\,.
\eae
In writing Eq.\,(\ref{masseNew}), Eq.\,(9) of \cite{Hofmann2017} and Eqs.\,(\ref{zerolam}), (\ref{rhovsrhogsT0})  
were used. Interactions between the monopole and 
deconfining SU(2) thermodynamics (caloron and anticaloron centers \cite{Hofmann2016} invoking 
energy transfer from quasi-particle excitations to the monopole which partially 
radiates this energy back into the Yang-Mills plasma) 
will effectively introduce spatial fluctuations of temperature $T$ about 
the equilibrium value $T_0$, but they won't change the energy content 
of the overall system. Solving Eq.\,(\ref{masseNew}) for $R_0$, we obtain the mean 
radial extent of the system as 
\eqb
\label{R_0New}
R_0=4.10\,H^{-1}_\infty(T_0)\,.
\eqe
To ensure that the use of infinite-volume thermodynamics, which has led to Eq.\,(\ref{R_0New}), is consistent we need to compare 
$R_0$ to the correlation length $|\phi|^{-1}(T_0)=\sqrt{2\pi T_0/\Lambda^3}$ 
of the thermal ground state \cite{HofmannBook2016}. Appealing to Eq.\,(\ref{Higgsasy}) and Eq.\,(\ref{YMscale}) below, one easily derives
\eqb
\label{phiinHinft}
|\phi|^{-1}(T_0)=H^{-1}_\infty(T_0)\sqrt{2\left(\frac{118.6}{12.96}\right)^3}\,.
\eqe
Thus 
\eqb
\label{R0phi}
R_0 |\phi|(T_0)=160.5\,,
\eqe
which justifies the use of infinite-volume thermodynamics in deriving Eq.\,(\ref{R_0New}). 

Comparing $R_0$ to the Compton wave length $\lambda_C=m_0^{-1}=2.43\times 10^{-12}\,$m, we have
\eqb
\label{CWL}
R_0\sim 53.14\,\lambda_C\sim 1.29\times 10^{-10}\,\mbox{m}\,.
\eqe
Thus, $R_0$ is 2.43 times larger than the Bohr radius $a_0$ which, in turn, is $4.10\times 12.96/2.43=21.87$ times 
larger than the monopole core radius $R_c=\frac{1}{e}H^{-1}_\infty(T_0)=2.43\times 10^{-12}\,$m$=\lambda_C$. 
It is remarkable that the relation between electron mass 
$m_0$ (that is, monopole-core breathing frequency $\omega_0$) and $H_\infty$, see Eq.\,(\ref{freutomass}), as found in 
\cite{FodorRacz2004}, is in agreement with $R_c^{-1}=m_0$ (or $R_c=\lambda_c$). Therefore, 
the electron is {\sl not} a point particle: Its charge is distributed over a spatial region whose radius 
matches the Compton wave length $\lambda_C$ but (quantum) moves within a much larger volume of radius 
$R_0\sim 10^{-10}\,$m of deconfining phase. Scattering experiments do not reveal an inner structure because {\sl quantum thermodynamics} within radius $R_0$ -- the state of maximum entropy -- is structureless. Moreover, the Yang-Mills scale $\Lambda$, which is determined from $H_\infty(T_0)=\pi T_0$ 
(monopole originated by dissociation of maximum-holonomy caloron) and Eq.\,(\ref{masseNew}), reads
\eqb
\label{YMscale}
\Lambda=\frac{1}{118.6}\,m_0\,.
\eqe
Therefore, $T_c=13.87/(2\pi\times 118.6)\,m_0=0.019\,m_0=9.49\,$keV. For a comparison, the demonstrator tokamak 
ITER is envisaged to operate at an average electron temperature of 8.8\,keV!   

The important work \cite{FodorRacz2004} points out that high-frequency 
oscillations in the profile function $w$ of the off-Cartan gauge fields, belonging to the perturbed monopole, develop 
expanding shells. If these shells are considered to penetrate the 
confining phase ($r>R_0$) such that the static Coulomb field of the monopole, arising from a temporal average 
over many cycles of the monopole core-vibration, can actually permeate this 
phase, then we may estimate the Coulomb-field correction $\Delta m_0$ to the ``thermodynamical" mass $m_0$ as
\eqb
\label{Coulomb} 
\Delta m_0=\int_{R_0}^{\infty} dr\,\frac{r^2}{r^4}=R_0^{-1}=\frac{1}{4.10} H_\infty(T_0)\ll 
12.96\,H_\infty=m_0\,\ \ \ \mbox{or} \ \ \ m_0\sim 53.14\,\Delta m_0\,.
\eqe
This is only a $\sim$1.9\,\% correction to $m_0$. Notice the strong conceptual difference to the model of the classical 
electron whose radius defines the entire rest mass $m_0$ via Coulomb self-energy.  

\noindent Finally, we would like to mention that, in assigning half a Bohr 
magneton to the magnetic moment carried by a single center-vortex loop, a $g$-factor of 
two naturally arises by the electron being composed of two such center-vortex 
loops by virtue of the region of self-intersection, see Fig.\,\ref{fig-1}. 

\section{Summary\label{Sum}}

The present article's purpose is a model of the electron along the lines proposed in \cite{Hofmann2017} with more precise inputs from SU(2) Yang-Mills thermodynamics and relying on the work in \cite{FodorRacz2004}. This model posits the electron 
to be represented by a one-fold self-intersecting spatial center-vortex loop immersed into the confining phase. 
The intersection region -- a blob of deconfining phase --, whose center-of-mass position essentially is a modulus 
of this solitonic configuration, contains a BPS monopole responding to perturbations issued by the 
surrounding thermodynamics. This interplay between non-linear monopole dynamics and the thermal plasma 
gives rise to a core-breathing mode of the monopole which runs an internal clock. The existence of such a localized vibration was 
foreseen by de Broglie already ninety years ago \cite{deBroglieDoc}. That this assertion, which is at the heart of the 
proposal of the electron's wave-particle duality, in turn leading to the development of wave mechanics 
by E. Schr\"odinger \cite{Schroedinger}, was more or less forgotten during the century to follow definitely is owed to the success of QM 
as a quantitative, physical theory. This is a truly remarkable state of affairs.  

The space external to the intersection region is subject to the 
confining phase which, however, is 
permeated by high-frequency expanding shells \cite{FodorRacz2004}, thus providing a way for the Coulomb 
field of the monopole to reach beyond the self-intersection region. We have shown that the radius of self-intersection 
region is comparable to atomic dimensions with the monopole core size matching the electron's Compton wave length. Coulomb self-energy turns out to be a small correction to the energy of the intersection region. That the bulk of 
the electron mass and its charge is represented by the "thermodynamics of an isolated particle" is an early and exceptionally deep 
insight by Louis de Broglie \cite{DeBroglieRev}. In the present work we have indicated that SU(2) Yang-Mills 
thermodynamics supports this idea.             

We have not discussed the complicated situation of a radiating electron, which interacts with external fields, and the associated problem of radiation reaction. Also, it is not yet clear in detail how the weak 
interactions are accomodated into our model although the presence of massive vector modes with a hierarchically higher 
mass is assured by the confining phase \cite{HofmannBook2016}. Hopefully, all this can be worked out in the near future. From the 
present treatment it is obvious that other charged leptons -- the muon and the $\tau$-lepton -- should 
admit a description identical to that of the electron: One considers accordingly larger Yang-Mills scales of the associated 
SU(2) gauge theories and introduces mixing of Cartan algebras.

\section*{Acknowledgments} 

We would like to acknowledge interesting 
and stimulating conversations with Steffen Hahn and Yaron Hadad. We
acknowledge the financial support of the Deutsche Forschungsgemeinschaft and Ruprecht-Karls-Universit\"at Heidelberg within the funding programme Open Access Publishing.

\end{document}